\documentstyle[twocolumn,epsf]{jpsj_sissa}
\newcommand{\eps}{\varepsilon}
\newcommand{\vphi}{\varphi}
\newcommand{\dis}{\displaystyle}
\def\runtitle{Quantum Effects in Resistance-shunted Josephson
Junctions}
\def\runauthor{Takeo {\sc Kato} and Masatoshi {\sc Imada}}
\title{Quantum Effects of Resistance-Shunted Josephson Junctions}

\author
{Takeo {\sc Kato} and Masatoshi {\sc Imada} }

\inst
{Institute for Solid State Physics, 
University of Tokyo,\\ 7-22-1 Roppongi, Minato-ku, Tokyo 106-8666}

\recdate{August 3, 1999}

\abst{
Thermodynamics and transport properties of 
resistance-shunted Josephson junctions are studied 
theoretically in the tight-binding limit 
$E_{\rm C}/E_{\rm J} \ll 1$,
where $E_{\rm C}$ and $E_{\rm J}$ are
a charging energy and a Josephson coupling energy
respectively. Based on a phenomenological
harmonic-oscillator model,
weak coupling region $K=R_{\rm Q}/R \ll 1$ is 
analytically studied, where $R$ and $R_{\rm Q}=h/(2e)^2$ 
are a shunted resistance and the quantum resistance.
In addition to the effective bandwidth $2 \hbar \Delta_{\rm eff}$, 
we find that this multi-level system genuinely has a novel crossover
at lower energy $K \hbar \Delta_{\rm eff}$ 
below which the density of states becomes strongly degenerate.
These two energy scales control the linear DC responses, 
optical responses, and nonlinear $I$-$V$ characteristics. 
The lower energy crossover indicates
the existence of a new class of 
strongly-correlated phenomena beyond the framework of
the Kondo problem.}

\kword
{Josephson junctions, dissipation, Coulomb blockade,
macroscopic quantum effects, superconductor-insulator transition}
\begin{document}
\sloppy
\maketitle
%
%
\section{Introduction}
\label{sec1}
Small resistance-shunted Josephson junctions are simplest systems
for the purpose of clarifying
various macroscopic quantum effects such as quantum
tunneling, quantum coherence, Coulomb blockade, 
etc.~\cite{Caldeira83,Leggett87,Zaikin90,Weiss93}
Though quantum properties of Josephson junctions
have been studied in many literatures,~\cite{Weiss93} 
thermodynamics and I-V characteristics 
have not been clarified except for
some special limits. In this paper, we study 
specific heat and optical linear responses 
in the weak damping region analytically. We also conjecture
the nonlinear I-V characteristics from the available results.

\begin{figure}[b]
  \hfil
  \epsfxsize=35mm
  \epsfbox{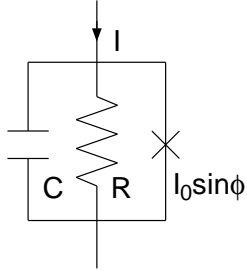}
  \hfil
  \caption{The equivalent circuits for a small Josephson junction}
  \label{circuit}
\end{figure}

Classical dynamics of the Josephson junctions has been 
studied based on the equivalent circuits shown in Fig.~\ref{circuit}.
The equation of motion is described as
\begin{equation}
C\dot V + \frac{V}{R} + I_0 \sin \phi = I,
\label{eq1}
\end{equation}
where $C$, $R$, $I_0$, $I$ are a capacitance, a shunted-resistance,
a Josephson critical current, and an external current respectively.
The dynamics of the phase difference $\phi$ generates a voltage
$V$ across the junction through the Josephson relation
\begin{equation}
V =\frac{\Phi_0}{2\pi} \dot \phi,
\label{eq2}
\end{equation}
where $\Phi_0=h/2e$ is a unit flux. From (\ref{eq1}) and (\ref{eq2}), 
the equation of motion is written as 
\begin{eqnarray}
& & C \ddot q + \frac{1}{R} \dot q + \frac{\partial U}{\partial q} =
0,  \label{eq3} \\
& & U(q) = - I q - E_{\rm J} \cos 
\left(2\pi \frac{q}{\Phi_0} \right),
\end{eqnarray}
where $q=\Phi_0 \phi / 2\pi$ and $E_{\rm J}= \Phi_0 I_0/2\pi$.
This equation can be interpreted as an equation of motion
of a dissipative particle with a coordinate $q$
in the washboard-like potential $U(q)$. The mobility $\mu$
of this dissipative particle is related to the I-V characteristics 
of the junction as
\begin{equation}
V =\mu I.
\end{equation}

The Josephson junction has two important energy scales:
one is the Josephson coupling energy $E_{\rm J}$, and the 
other is the charging energy $E_{\rm C}=e^2/2C$.
The ratio $E_{\rm J}/E_{\rm C}$
controls the quantum effects in Josephson junctions.
When the area of the junction $S$ becomes small,
the charging energy $E_{\rm C}(\propto 1/S)$ cannot be neglected
compared with $E_{\rm J}(\propto S)$. Then,
the system cannot be described by the classical equation.
To study dissipative quantum systems,
a phenomenological harmonic-oscillator 
model is introduced:~\cite{Caldeira83,Leggett87}
\begin{eqnarray}
H &=& H_0 + \sum_{j} \left\{ \frac{p_j^2}{2m_j}
+ \frac12 m_j \omega_j^2 
\left(x_j - \frac{c_j}{m_j \omega_j^2} q \right)^2 \right\}, 
\label{CLHam} \\
H_0 &=& \frac{Q^2}{2C} + U(q), 
\end{eqnarray}
where $Q$ is a pseudo-momentum conjugate to $q$, which
corresponds to a charge of the junction.
The harmonic oscillators represent an environment which
generates the damping term. When the spectral function 
is taken as
\begin{eqnarray}
J(\omega) &\equiv& \frac{\pi}{2} \sum_j \frac{c_j^2}{m_j \omega_j}
\delta (\omega-\omega_j) \\
&=& \frac{\omega}{R} e^{-\omega/\omega_{\rm c}},
\end{eqnarray}
the equation of motion (\ref{eq3}) is reproduced in the
classical limit. Here, $\omega_{\rm c}$ is a cut-off
frequency.

The ground state of this model 
has been discussed by using perturbative
renormalization analysis and duality mapping.~\cite{Fisher85}
The dimensionless damping strength defined by
$K = R_{\rm Q}/R$ controls the ground-state properties,
where $R_{\rm Q} = h/4e^2 (=6.5 [{\rm k}\Omega])$ is the quantum
resistance. For $K<1$, the particle is extended in the real space.
In this case, superconductivity of the Josephson junction is destroyed
by large phase fluctuations, and the junction 
behaves as an insulator.
For $K>1$, the particle is localized in a well, and the 
superconductivity of the Josephson junction is retained.
Thus, it is predicted that 
there exists the superconductor-insulator transition at $K=1$.
These results have been supported on a qualitative level
by experiments.~\cite{Yagi97,Penttila99}

The $I$-$V$ characteristics of the resistance-shunted junction
have been studied in the
literatures.~\cite{Zaikin90,Weiss93,Eckern87,Zwerger87,Korshunov87,Weiss85,Chen89,Weiss91,Sassetti92}
Generally, it is difficult to study the dissipative model for 
arbitrary ratio $E_{\rm C}/E_{\rm J}$. 
In this paper, we restrict ourselves
to the tight-binding limit $E_{\rm C}/E_{\rm J} \ll 1$, where
the system Hamiltonian $H_0$ is well described by
a one-dimensional tight-binding model
\begin{equation}
  H_0 = - \hbar \Delta \sum_l \left( c_{l+1}^{\dagger} c_l
  + c_l^{\dagger} c_{l+1} \right),
  \label{tbHam}
\end{equation}
in the absence of the external current.
Here, $\Delta$ is a hopping amplitude, and $c_l$ $(c_l^{\dagger})$
is an annihilation (creation) operator at the site $l$.
Even in this limit, the properties of the junction
in the presence of the dissipation have been
studied only for special regions in the whole parameter space. 
So far, there are two regions which allows analytical 
calculations: one is 
the high-temperature or/and strong-damping limit,~\cite{Weiss85}
and the other is the $K=1/2$ solvable line.~\cite{Weiss91}

Recently, a formulation for $K\ll 1$ has been 
proposed by the authors.~\cite{Kato98}
Based on this formulation, we can calculate
thermodynamics and linear responses at any temperature.
This calculation reveals the $K$-dependence 
of the low-temperature properties, which have not been 
studied so far. 
The calculation for the weak damping region reveals
a novel feature. In the absence of dissipation,
only the bandwidth $2 \hbar \Delta$ determines
the characteristic energy scale of the junction. 
Additionally,
a novel characteristic energy scale $K \hbar \Delta$ appears in the
thermodynamics and transport properties. This feature can be 
observed most clearly in the weak damping region $K \ll 1$,
and cannot be observed clearly in the $K=1/2$ solvable case.

In Ref.~\citen{Kato98}, the authors have treated the
dissipative tight-binding model as an effective
model of a dissipative carrier in solids, and
have not studied applications to Josephson junctions.
In this paper, by using the 
results of Ref.~\citen{Kato98} and other available references,
the aspects of thermodynamics and $I$-$V$ characteristics of
Josephson junctions are presented in detail. 
This paper has two purposes: (1) we report
comprehensive studies on the existence of the 
two different energy scales (especially for experimentalists),
and (2) we analyze and conjecture the junction properties
for general values of $K$ beyond the solvable cases.

This paper is organized as follows. The formulation of
the weak damping region is summarized briefly in Sec.~\ref{sec2}.
The thermodynamics and $I$-$V$ characteristics are
discussed in Sec.~\ref{sec3}. 
In Sec.~\ref{sec4}, the relation to the Kondo problem
is discussed, and finally in Sec.~\ref{sec5}
the results are summarized. 

\section{Formulation}
\label{sec2}

The partition function is formulated
from (\ref{CLHam}) and (\ref{tbHam}) 
in terms of the path integrals as~\cite{Kato98} 
\begin{equation}
  Z = \sum_{m=0}^{\infty} \frac{\Delta^{2m}}{2m!}
  \sum_{\{\xi_l\}'} \prod_{n=1}^{2m} \int_0^{\hbar \beta}
  {\rm d}\tau_n
  \exp\left[\sum_{k<l}^{2m} \xi_k \xi_l \phi(\tau_l -
  \tau_k) \right] ,
  \label{Z}
\end{equation}
where $\beta = 1/k_{\rm B}T$, $\xi_l=\pm 1$, 
and the prime in $\{\xi_l\}'$ 
denotes the summation in accordance with
the constraint $\sum_l \xi_l = 0$.
The function $\phi(\tau)$ is defined by
\begin{eqnarray}
  \phi(\tau) &=& 
  \frac{\Phi_0^2}{\pi\hbar} \int_0^{\infty} {\rm d}\omega
  \frac{J(\omega)}{\omega^2}
  \Biggl( \coth(\hbar \beta \omega/2) \Biggr. \nonumber \\
  & & - \Biggl. \frac
  {\cosh \left[ \omega(\hbar \beta/2 - |\tau|)\right]}
  {\sinh(\hbar \beta\omega/2)}
  \Biggr).
\end{eqnarray}
The expression (\ref{Z}) may be interpreted as a statistical
model of classical interacting particles with
the potential $\phi(\tau)$.

The weak coupling theory can be constructed by applying
the Debye-H\"uckel approximation to this 
classical partition function.~\cite{Kato98} 
It has not been proved that this approximation
is valid for the present model.
It, however, can be shown that this approximation
reproduces correct results at least in the high-temperature region.
It also reproduces the correct results in
the continuum limit ($K \rightarrow 0$, $\Delta \rightarrow \infty$, 
$K\Delta = {\rm const.}$). Hence, we expect that this approximation
gives correct qualitative features.
The validity of the Debye-H\"uckel approximation is discussed
in Appendix.

By the Debye-H\"uckel approximation, the partition function is
calculated as~\cite{Kato98}
\begin{eqnarray}
& & Z = \int_0^{2\pi} \frac{{\rm d}\theta}{2\pi} {\rm e}^{U(n)},
\label{partition} \\
& & U(n) = n + \log\Gamma(Kn+1) - Kn\psi(Kn+1),
\label{Unexp} \\
& & Q(n) = K(\psi(Kn+1)+\bar{\gamma}),
\label{Qnexp} \\
& & 2\beta \hbar \Delta
\left(\frac{\hbar \beta\omega_{\rm c}}{2\pi}\right)^{-K}
\! \! \cos\theta = n {\rm e}^{-Q(n)},
\label{nthetaexp}
\end{eqnarray}
where $\bar{\gamma}$ is the Euler constant, $\Gamma(x)$ 
is the gamma function, and $\psi(x)$ is the polygamma function.
The specific heat is directly calculated from the partition 
function. The density of states $D(\omega)$ is 
also calculated from $Z(\beta)$ as
\begin{equation}
Z(\beta) = \int_{\omega_0}^{\infty} {\rm d}\omega 
D(\omega) e^{-\hbar \beta \omega},
\label{Laplace}
\end{equation}
where $\hbar \omega_0$ is the ground state energy.

We formulate the optical mobility $\mu(\omega)$
by the Kubo formula. The exact formal expression for
$\mu(\omega)$ is given by~\cite{Kato98}
\begin{eqnarray}
\mu(\omega) &=& 
\frac{\Phi_0^2}{\hbar \omega} \,
{\rm Im} \int_0^{\infty} {\rm d}t \, \Lambda(t)
e^{{\rm i}\omega t}, \\
\Lambda(t) &=& -2 \, {\rm Im} \tilde{\Lambda} 
(\tau\rightarrow {\rm i}t), \\
\tilde{\Lambda}(\tau) &=& \frac{\Delta^2}{Z}
\sum_{m=0}^{\infty} \sum_{\{\xi_l,\sigma,\sigma'\}'}
(-\sigma \sigma') \frac{\Delta^{2m}}{2m!} \nonumber \\
&\times& \prod_{l=1}^{2m}
\int_0^{\hbar \beta} {\rm d}\tau_l \exp [ S(\{\tau_l\},\tau) ], \\
S(\{\tau_l\},\tau) &=& \sum_{k<l}^{2m} \xi_k \xi_l \phi(\tau_l-\tau_k)
+ \sum_{l=1}^{2m} \sigma \xi_l \phi(\tau_l) \nonumber \\
&+& \sum_{l=1}^{2m} \sigma' \xi_l \phi(\tau_l - \tau)
+ \sigma \sigma' \phi(\tau),
\end{eqnarray}
where $\xi_l, \sigma, \sigma' = \pm 1$,
and the prime
in $\{\xi_l, \sigma, \sigma'\}'$ denotes the summation
in accordance with the constraint $\sum_l \xi_l + \sigma + \sigma' 
= 0$. Here, the current-current correlation function 
$\tilde{\Lambda}(\tau)$
can be interpreted as a classical
partition function with fixed charges at $0,\tau$.
Hence, the Debye-H\"uckel approximation is also
applicable in the calculation of $\tilde{\Lambda}(\tau)$
in the weak damping region $K \ll 1$. As a result,
the real-time 
correlation function $\Lambda(t)$ is calculated as
\begin{eqnarray}
\Lambda(t) &=& \frac{1}{Z} \int_0^{2\pi} \frac{{\rm d}\theta}{2\pi}
\frac{n^2}{\hbar^2 \beta^2 \cos^2 \theta}e^{U(\theta)} 
\nonumber \\
&\times& \left[ e^{-S(t;\theta)} + e^{S(t;\theta)} \cos 2\theta \right]
\sin R(t;\theta),
\end{eqnarray}
where $R(t;\theta)$ and $S(t;\theta)$ are defined as
\begin{eqnarray}
R(t;\theta) &=& \pi K e^{-\gamma(\theta) t}, \\
S(t;\theta) &=& \frac{\pi}{\hbar \beta \gamma(\theta)} 
-\frac{\pi}{2}
\cot \frac{\hbar \beta \gamma(\theta)}{2} e^{-\gamma(\theta)t}
\nonumber \\
&-& \sum_{k=1}^{\infty} 
\frac{k\exp(-2\pi k t/\hbar \beta)}
{k^2-(\hbar \beta \gamma(\theta)/2\pi)^2}, \\
\gamma(\theta) &=& 2\pi K n(\theta)/\hbar \beta.
\end{eqnarray}
Thus, the linear mobility can be calculated at any frequencies.

\section{Results}
\label{sec3}

In this section, we show results mainly for the weak damping
region $K \ll 1$ based on the formulation of Sec.~\ref{sec2}. 
In this paper, we stress the fact that the two 
different energy scales, $K \hbar \Delta_{\rm eff}$
and $\hbar \Delta_{\rm eff}$ control the junction properties.
We expect that this new viewpoint is helpful to understand the 
junction properties for general values of $K$ and $E_{\rm C}/
E_{\rm J}$. 

In addition to the weak damping region $K \ll 1$,
there exist some regions where analytical result can be obtained:
the $K=1/2$ exact solution~\cite{Weiss91}
and incoherent hopping regime.~\cite{Leggett87,Weiss85}
From these results, the thermodynamics and $I$-$V$ 
characteristics are conjectured 
beyond the weak damping region. 

In the following subsections, 
we discuss the thermodynamics (Sec.~\ref{sec30}),
the linear DC mobility (Sec.~\ref{sec31}), the linear
optical mobility 
(Sec.~\ref{sec32}), and the nonlinear DC
mobility (Sec.~\ref{sec33}).
Finally in Sec.~\ref{sec34}, the $I$-$V$ characteristics
are derived from a simplified model, and is compared with
our results.

\subsection{Thermodynamics}
\label{sec30}

The specific heat of the dissipative particle has been calculated 
at any temperature in Ref.~\citen{Kato98}. 
The low- and high-temperature form of the specific heat $C$ 
is calculated for $K \ll 1$ as
\begin{equation}
C/k_{\rm B} = \left\{ \begin{array}{ll}
\dis{4 \pi^2 \left(\frac{2\pi k_{\rm B}T}
{\hbar \Delta_{\rm eff}}\right)^{2K-2}},
& ( k_{\rm B}T \gg \hbar \Delta_{\rm eff} ), \\
\dis{\frac{k_{\rm B}T}
{12K c \hbar \Delta_{\rm eff}}}, & 
( k_{\rm B} T \ll \hbar \Delta_{\rm eff}), 
\end{array} \right.
\end{equation}
where the effective hopping amplitude is defined by 
$\Delta_{\rm eff} = \Delta (\Delta/\omega_{\rm c})^{K/(1-K)}$.
The constant $c = (4\pi K e^{\bar{\gamma}})^{K/(1-K)}$ 
is close to unity for $K \ll 1$.

In this paper, we focus on strong modification of
the density of states caused by dissipation.
The density of low-energy states is obtained
from the low-temperature expansion of the partition function
as~\cite{Kato98}
\begin{equation}
D(\omega) = \sqrt{K} \delta(\omega) + \frac{1}{12 \sqrt{K} W}
+ {\cal O}(\omega),
\label{weightD}
\end{equation}
where $W=2 c \Delta_{\rm eff}$ is the effective bandwidth,
and the energy is shifted so that the ground-state energy becomes
zero. This shape of $D(\omega)$ is 
quite different from the one in the dissipationless case
where $D(\omega)\propto \omega^{-1/2}$.

\begin{figure}[tb]
  \hfil
  \epsfxsize=70mm
  \epsfbox{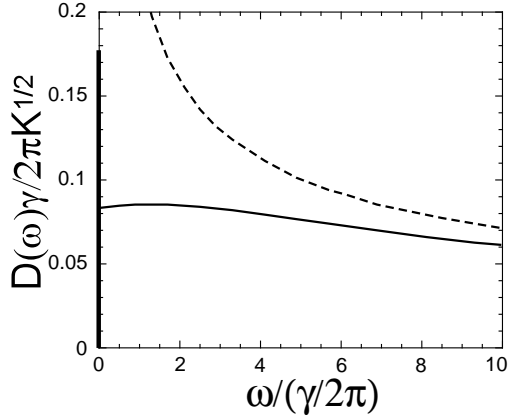}
  \hfil
  \caption{The density of low-energy states $D(\omega)$ in the continuum
    limit for the dissipative case (the solid curve) and the 
    dissipationless case (the dashed curve). The density of
    states for the dissipative case 
    has a delta function at $\omega = 0$, which is
    shown schematically by the thick line.}
  \label{DOS}
\end{figure}

We can obtain the whole shape of $D(\omega)$ analytically
in the continuum limit: $K \rightarrow 0$
and $\Delta \rightarrow \infty$ with $K \Delta$ fixed.
In this limit, the particle can be treated as
a dissipative Brownian particle with a damping coefficient
$\gamma = 4\pi K\Delta$, which is the only energy 
scale of the system.~\cite{Kato98,footnote} In the continuum limit, 
the integral (\ref{partition}) is determined by the contribution
around $\theta = 0$, and can be evaluated by 
the saddle point method. Thus, the partition function is
calculated as
\begin{eqnarray}
\log Z &=& 2 \hbar \beta \Delta + \frac12 \log (2\pi K)  
+ \log \Gamma
\bigl(\frac{\hbar \beta \gamma}{2\pi} + 1\bigr)
\nonumber \\
&-& \bigl(\frac{\hbar \beta \gamma}{2\pi} + 
\frac12 \bigr)\log 
\frac{\hbar \beta \gamma}{2\pi} .
\label{contZ}
\end{eqnarray}
From this expression, the density of states
$D(\omega)$ can be obtained numerically by the inverse Laplace 
transform (\ref{Laplace}). The result takes the form
\begin{equation}
  D(\omega) = \sqrt{K} \delta(\omega) + \tilde{D}(\omega).
\end{equation}
The excited-state part $\tilde{D}(\omega)$ is shown in Fig.~\ref{DOS} by 
the solid curve, while the dissipationless case is
shown by the dashed curve. 
The delta function at $\omega = 0$ is also
schematically drawn by the thick line in Fig.~\ref{DOS}. 

\begin{figure}[tb]
  \hfil
  \epsfxsize=60mm
  \epsfbox{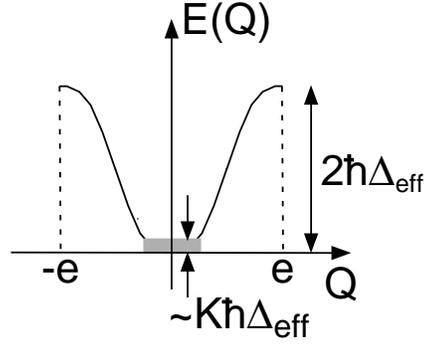}
  \hfil
  \caption{The dispersion relation of the junction.
    Besides the energy scale of the band width $2 \hbar
    \Delta_{\rm eff}$,
    there appears a new characteristic energy 
    scale $\hbar \gamma \sim K \hbar \Delta_{\rm eff}$ below which the
    dispersion is strongly modified in the gray region.}    
  \label{disp}
\end{figure}

If the dissipation effect appeared only in the suppression
of the bare tunneling matrix $\Delta$ to $\Delta_{\rm eff}$,
the band dispersion would be given as 
\begin{equation}
  E(Q) = 2 \hbar \Delta_{\rm eff} 
  (1-\cos (\pi Q/e)).
  \label{dispeq}
\end{equation}
This dispersion, however, does not reproduce the present result
of $D(\omega)$. In particular, the delta function
in $D(\omega)$ at $\omega = 0$ indicates that the 
low-energy states are degenerate around the band bottom.
The qualitative feature of the dispersion is shown 
in Fig.~\ref{disp}, where the dispersion relation
of relevant excitation is strongly
modified in the gray region ($E < K \hbar \Delta_{\rm eff}$).
For the moment, it is not yet confirmed whether this dispersion
is directly related to the single particle
dispersion relation because
single-particle Green's function has not been obtained yet because of 
difficulty in the analytical calculation even for $K \ll 1$.

As seen in Fig.~\ref{disp},
the dissipative system of $K \ll 1$ has 
two different energy scales: the bandwidth $\hbar \Delta_{\rm eff}$
and the damping energy $\hbar \gamma \sim K \hbar \Delta_{\rm eff}$.
These two energy scales appear also in
the $I$-$V$ characteristics as shown in the following subsections.
Note that the existence of these two energy scales is not
an artifact of the Debye-H\"uckel approximation. Actually,
the nontrivial energy scale $\hbar \gamma$ appears in
the continuum limit without the Debye-H\"uckel
approximation by mapping the model to the Brownian motion problem,
and the partition function (\ref{contZ}) can be reproduced
except for the divergent part $\log K$. (See also Appendix.)

As seen in (\ref{weightD}),
the weight of the delta function in $D(\omega)$ increases
as $\sqrt{K}$ for $K \ll 1$. Hence,
we expect that the macroscopic degeneracy at $\omega = 0$
persists for general values of $K$, and that the weight of the 
delta function in $D(\omega)$ increases as $K$ is enlarged.
This expectation
gives a new aspect on the superconductor-insulator
transition at $K=1$. As $K$ approaches unity, the effective
bandwidth 
$\hbar \Delta_{\rm eff} = \hbar\Delta(\Delta/\omega_c)^{K/(1-K)}$ 
is suppressed to be zero,~\cite{Leggett87}
and all the states become degenerate.
Our result shows that a partial degeneracy remains
`even for $K<1$'. 

This degeneracy in the density of states
may be observed by the time-evolution measurement of the voltage, 
when the external current is large enough to cause 
the Bloch oscillation. If we consider a simplified model
as discussed in Sec.~\ref{sec34}, the flat dispersion 
generates no voltage. Hence, the time-dependence of the 
voltage may show the region of zero voltage for some interval.

\subsection{Linear DC responses}
\label{sec31}

\begin{figure}[tb]
  \hfil
  \epsfxsize=65mm
  \epsfbox{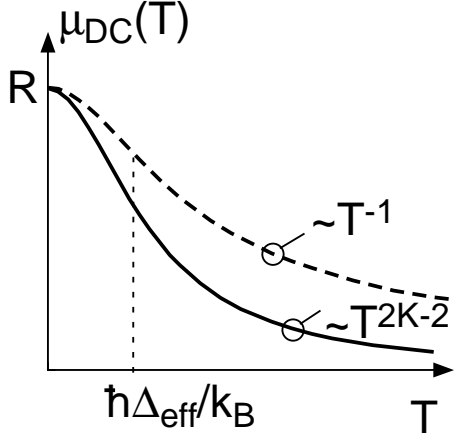}
  \hfil
  \caption{The temperature-dependences of the DC mobility
    $\mu_{\rm DC}(T)$ are shown schematically
    by the solid curve for $K \ll 1$,  
    and by the dashed curve for $K=1/2$.} 
  \label{Fig1}
\end{figure}

The low- and high-temperature behavior
of the DC mobility $\mu_{\rm DC}(T)$ can be
obtained within the linear response theory
for the weak damping region $K \ll 1$ as~\cite{Kato98}
\begin{equation}
\mu_{\rm DC}(T) = \left\{ 
\begin{array}{l}
\dis{ R \left( 1- \frac18 \left( \frac{k_{\rm B}T}{ 
c \hbar \Delta_{\rm eff}} \right)^2
+ {\cal O}(T^4) \right) }, \\
\hspace{10mm} 
(k_{\rm B}T \ll \hbar \Delta_{\rm eff}), \\
\dis{ 8\pi^2 R \left(\frac{
\hbar \Delta_{\rm eff}}{2\pi k_{\rm B}T} \right)^{2-2K}}, \\
\hspace{10mm}
(k_{\rm B}T \gg \hbar \Delta_{\rm eff}).
\end{array} \right.
\label{weakDC}
\end{equation}
The qualitative features of $\mu_{\rm DC}(T)$ for $K\ll 1$
obtained by connecting high- and low-temperature
behavior are shown in Fig.~\ref{Fig1} by the solid curve.
At high temperatures, the mobility
increases as the temperature decreases, while at
low temperatures, the mobility saturates to $R$.
The crossover temperature appears at
$T \sim \hbar \Delta_{\rm eff}/k_{\rm B}$.

We expect that this feature of $\mu_{\rm DC}(T)$
is universal for $0<K<1$. In fact, the feature of
(\ref{weakDC}) is valid for the special case $K=1/2$, which 
allows analytical treatment.~\cite{Weiss91} 
In this case, $\mu_{\rm DC}(T)$ is calculated 
exactly as
\begin{eqnarray}
\mu_{\rm DC}(T) &=& 
R \, \frac{\hbar\Delta_{\rm eff}}{2 k_{\rm B}T} 
\psi'(\hbar \Delta_{\rm eff}/2k_{\rm B}T + 1/2), \\
&=& \left\{ \begin{array}{l}
R \displaystyle{\left( 1 - \frac13 \left(\frac{k_{\rm B}T}
{\hbar \Delta_{\rm eff}} \right)^2 \right)}, \\
\hspace{10mm} (k_{\rm B}T \ll \hbar 
\Delta_{\rm eff}), \\
R \displaystyle{\frac{\pi^2 \hbar \Delta_{\rm eff}}
{4 k_{\rm B} T}}, \\
\hspace{10mm}(k_{\rm B}T \gg \hbar \Delta_{\rm eff}), 
\label{halfDC} 
\end{array} \right.
\end{eqnarray}
where $\psi'(x)$ is the digamma function.
The qualitative behavior of $\mu_{\rm DC}(T)$
is shown in Fig.~\ref{Fig1} by the dashed curve.
This result of the $K=1/2$ case is consistent with
our result (\ref{weakDC}).

These behaviors of $\mu_{\rm DC}(T)$ have been
interpreted physically as follows.~\cite{Fisher85} 
At high temperatures,
the phase $\phi$ slips by $2\pi$ thermally. This 
phase slip occurs incoherently with a temperature-dependent
tunneling rate determined essentially by 
the incoherent hopping in the two-level systems.
As the temperature decreases, the phase-slip picture 
becomes inadequate, and the quantum coherence of the system
appears. The wave function begins
to extend in the $\phi$ space. Because of the large
phase fluctuation,
the Josephson junction behaves as an insulator,
where Cooper pairs cannot tunnel through the junction.
In this situation, the whole part of the current flows through the
shunted resistance $R$, and the voltage
follows the ohmic law $V = IR$, 
which corresponds to $\mu_{\rm DC} = R$. 

\subsection{Linear optical responses}
\label{sec32}

\begin{figure}[tb]
  \hfil
  \epsfxsize=65mm
  \epsfbox{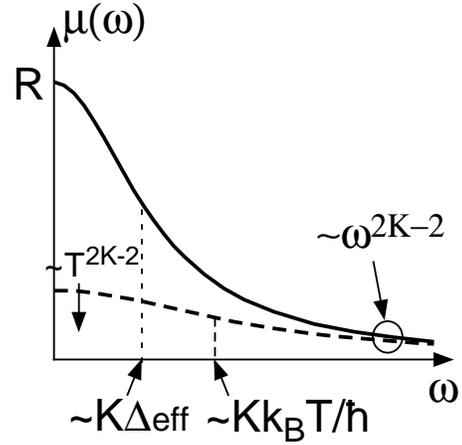}
  \hfil
  \caption{The optical mobility $\mu(\omega;T)$ for $K \ll 1$.
    The zero-temperature result is shown by
    the solid curve, 
    and the high-temperature result 
    by the dashed curve.}
  \label{Fig2}
\end{figure}

In this subsection, we calculate
the optical mobility $\mu(\omega;T)$
within the linear response theory, which
can be interpreted as the optical responses
of the Josephson junctions.

At zero temperature, the optical mobility can
be calculated analytically for $K \ll 1$
at any frequency, and the details of the results
have been given in Ref.~\citen{Kato98}.
Here, we discuss qualitative features of $\mu(\omega)$
from the analytical forms of the limiting regions:
\begin{equation}
\mu(\omega;T=0) = \left\{
\begin{array}{ll}
\dis{\frac{R e^{-2K\bar{\gamma}}}{2\Gamma(2K+1)} \left(
\frac{\gamma}{\omega}
\right)^{2-2K}}, & (\omega \gg \omega^*), \\
R \dis{\frac{\gamma^2}{\gamma^2 + \omega^2}}, 
& (\omega \ll \omega^*),
\end{array} \right. 
\label{mobcross2}
\end{equation}
where $\bar{\gamma}$ is the Euler constant.
The qualitative features of $\mu(\omega)$ 
are shown in Fig.~\ref{Fig2} by the solid curve. 
The damping frequency is given as
$\gamma = 4\pi K c \Delta_{\rm eff}$.
Thus, the characteristic energy
scale in $\mu(\omega)$ appears at
$\hbar \gamma \sim K \hbar\Delta_{\rm eff}$,
which is the same order as the energy scale
observed in the modification of the dispersion.
The crossover frequency $\omega^*$ in (\ref{mobcross2})
is evaluated as $\omega^* \sim 2^{1/K} \gamma$.
When we take $K \rightarrow 0$ and $\Delta \rightarrow
\infty$ with $K \Delta$ fixed, the optical responses
take exact Lorentzian forms ($\omega^* \rightarrow \infty$),
which are obtained also by considering
the Brownian motion.~\cite{Kato98}

As $K$ increases, it is expected that
the crossover frequency $\omega^*$ in (\ref{mobcross2}) decreases, 
and the deviation from Lorentzian form becomes clear.
At the same time,
the effective hopping amplitude $\Delta_{\rm eff}$ is suppressed.
Particularly, in the limit $K \rightarrow 1$, 
the amplitude $\Delta_{\rm eff}$ vanishes
and the characteristic frequency $\gamma\sim K\Delta_{\rm eff}$ 
in (\ref{mobcross2}) also vanishes.
For $K >1 $, the dissipative
particle moves only by the incoherent hopping, and
the optical mobility takes a different form~\cite{Weiss85}
\begin{equation}
\mu(\omega;T=0) = R \frac{8\pi^2 K^2 \Delta^2}
{\Gamma(2K+1)\omega_{\rm c}^2} \left(\frac{\omega}{\omega_{\rm c}}
\right)^{2K-2}.
\end{equation} 

The finite-temperature calculation of $\mu(\omega;T)$
is difficult even for $K \ll 1$. For a qualitative discussion, 
however, it is sufficient to consider at
high-temperature $k_{\rm B} T \gg \hbar \Delta_{\rm eff}$, where
the optical mobility is calculated for arbitrary values of $K$ 
by assuming the incoherent hopping of the dissipative
particle. The optical mobility is obtained for $0<K<1$
as~\cite{Weiss85}
\begin{eqnarray}
\mu(\omega;T) &=& R \, 4 \pi K \Delta_{\rm eff} 
\left(\frac{\hbar \Delta_{\rm eff}}{2\pi k_{\rm B}T}\right)^{2-2K}
\nonumber \\ &\times& 
\frac{|\Gamma(K+{\rm i}\hbar\omega/2\pi k_{\rm B}T)|^2}{\Gamma(2K)}
\frac{\sinh\hbar \omega/2 k_{\rm B}T}{\omega},
\label{mobcross3} \\
&=& \left\{ \begin{array}{l}
R \, \dis{\frac{8 \pi^2 K^2}{\Gamma(2K+1)}
 \left(\frac{\Delta_{\rm eff}}{\omega}\right)^{2-2K}}, \\
\hspace{10mm} (\omega \gg K k_{\rm B}T/\hbar), \\
R \, \dis{\frac{8 \pi^2 K^2 \Gamma(K)^2}{\Gamma(2K+1)}
\left(\frac{\hbar \Delta_{\rm eff}}{2\pi
k_{\rm B} T}\right)^{2-2K}},
\\ \hspace{10mm} (\omega \ll K k_{\rm B}T/\hbar).
\end{array} \right.
\end{eqnarray}
The qualitative behavior is shown by the dashed curve in Fig.~\ref{Fig2}.
The high-frequency side is temperature-independent, and agrees with
the zero-temperature result in the weak damping region $K \ll 1$.
This shows that the high frequency side is well described by
the incoherent hopping of the dissipative particle even at zero 
temperature. The low-frequency side is temperature-dependent,
and behaves as $T^{2K-2}$. The crossover frequency is
given as $\omega \sim K k_{\rm B} T/\hbar$.

When the temperature decreases, 
the low-frequency temperature-dependent part in (\ref{mobcross3})
becomes inadequate below a crossover temperature.
This crossover temperature is estimated for $K \ll 1$ as 
$T \sim \hbar \Delta_{\rm eff}/k_{\rm B}$ where the characteristic
frequency in (\ref{mobcross3})
takes the same order ($\gamma = 4\pi K\Delta_{\rm eff}
\sim K k_{\rm B}T/\hbar$) as that seen in (\ref{mobcross2}). 
This result is consistent with 
that of the DC mobility, where the 
crossover temperature is given also as 
$T \sim \hbar \Delta_{\rm eff}/k_{\rm B}$.

\subsection{Nonlinear responses}
\label{sec33}

Theoretically, it is a difficult task to study
nonlinear responses beyond the linear response 
theory, because nonequilibrium states must be treated properly.
In this paper, general properties of 
the $I$-$V$ characteristics are inferred
from the results obtained in some limits.

\begin{figure}[tb]
  \hfil
  \epsfxsize=65mm
  \epsfbox{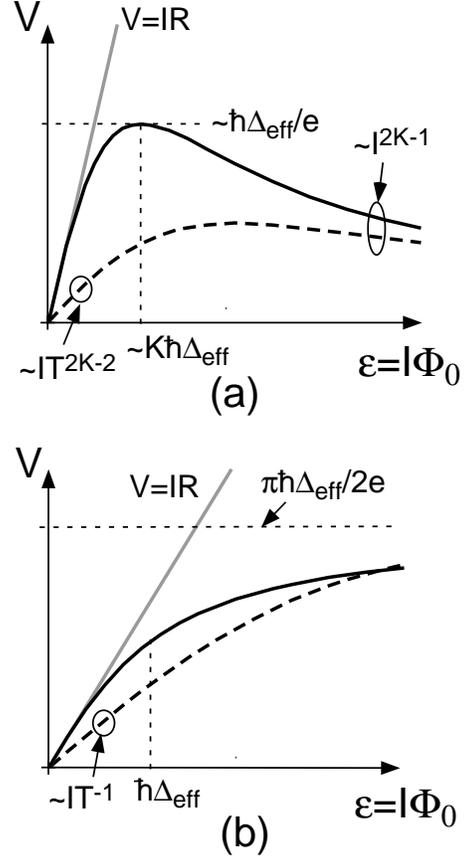}
  \hfil
  \caption{(a) The obtained $I$-$V$ characteristics for $0<K<1$, 
    and (b) the exact solution at $K=1/2$.
    The zero-temperature results are shown by the solid curves,
    and the high-temperature results by the dashed curves.
    The Ohmic responses $V=IR$ are also shown by gray lines
    for references.}
  \label{Fig3}
\end{figure}

At zero temperature, the nonlinear mobility 
$\mu_{\rm NL} = V/I$ can be calculated for large $I$ as~\cite{Weiss85}
\begin{eqnarray}
\mu_{\rm NL}(\eps;T=0) = R \frac{8 \pi^2 K^2}{\Gamma(2K+1)}
 \left(\frac{\hbar\Delta_{\rm eff}}{\eps}\right)^{2-2K},
\end{eqnarray}  
where $\eps=I \Phi_0$ is the energy difference between 
neighboring wells in the periodic potential $U(q)$.
For large $I$, the voltage $V = \mu_{\rm NL} I$ is proportional to
$I^{2K-1}$. On the other hand for small $I$, the mobility
$\mu_{\rm NL}$ 
is expected to be $R$ \ for $0< K < 1$ from the renormalization
group analysis.~\cite{Fisher85} By connecting these two limiting
cases smoothly, the $I$-$V$ characteristics 
are derived as in Fig.~\ref{Fig3}~(a) by the solid curve. 

The crossover current is determined by the cross point of 
the results of the two limiting cases. Particularly for $K\ll 1$, the
crossover current is $\eps \sim K\hbar \Delta_{\rm eff}$,
while the characteristic voltage is
$V \sim IR \sim \hbar \Delta_{\rm eff}/e$. Thus, also in the $I$-$V$
characteristics, two different energy scales, 
$\hbar \Delta_{\rm eff}$ and $K \hbar \Delta_{\rm eff}$ 
appear clearly.

The high-temperature limit $k_{\rm B}T 
\gg \hbar \Delta_{\rm eff}$ can also
be calculated analytically for $0 < K < 1$ as
\begin{eqnarray}
\mu_{\rm NL}(\eps;T) &=& R \, 4 \pi K \hbar \Delta_{\rm eff} \left(
\frac{\hbar \Delta_{\rm eff}}{2\pi k_{\rm B}T}\right)^{1-2K}
\nonumber \\
&\times& \frac{|\Gamma(K+{\rm i}\eps/2 \pi k_{\rm B}T)|^2}{\Gamma(2K)}
\frac{\sinh(\eps/k_{\rm B}T)}{\eps} \\
&=& \left\{ \begin{array}{l}
R \, \dis{\frac{8 \pi^2 K^2}{\Gamma(2K+1)}
\left(\frac{\hbar \Delta_{\rm eff}}{\eps}\right)^{2-2K}}, 
\\ \hspace{10mm} (\eps \gg k_{\rm B}T), \\
R \, \dis{\frac{8 \pi^2 K^2 \Gamma(K)^2}{\Gamma(2K+1)}
\left(\frac{\hbar \Delta_{\rm eff}}{2\pi
k_{\rm B} T}\right)^{2-2K}},
\\ \hspace{10mm} (\eps \ll k_{\rm B}T).
\end{array} \right.
\end{eqnarray}
The schematic $I$-$V$ characteristics
are shown also in Fig.~\ref{Fig3}~(a) by the dashed curve.
At large current ($\eps = \Phi_0 I \gg k_{\rm B}T$), the 
mobility is temperature-independent, and behaves as
$I^{2K-2}$, while at small current ($\eps \ll k_{\rm B}T$)
it becomes temperature-dependent. 
Note that in this regime
the nonlinear DC mobility is obtained from the linear optical
conductivity as $\mu_{\rm NL}(\eps;T)
=\mu(\omega\rightarrow \eps/\hbar ;T)$.

We expect that the obtained $I$-$V$ characteristics
are valid for any values of $K$. This conjecture is easily checked
by considering the exact solution of the
$K=1/2$ case given by~\cite{Weiss91}
\begin{eqnarray}
\mu_{\rm NL}(\eps;T)
&=& R \, \frac{4 \pi \hbar \Delta_{\rm eff}}{\eps} {\rm Im}
\, \psi\left(\frac12 + \frac{2\hbar \Delta_{\rm eff}}{k_{\rm B}T}
+ {\rm i} \frac{\eps}{2\pi k_{\rm B}T} \right) \nonumber \\
&=& \left\{ \begin{array}{l}
\dis{R \,\frac{2\pi^2 \hbar \Delta_{\rm eff}}{\eps} \tanh 
\left(\frac{\eps}{2k_{\rm B}T}\right) },
\\ \hspace{10mm} (k_{\rm B}T \gg \hbar \Delta_{\rm eff}), \\
\dis{R \,\frac{4\pi \hbar \Delta_{\rm eff}}{\eps}
\arctan \left(\frac{\eps}{4\pi \hbar \Delta_{\rm eff}}\right)},
\\ \hspace{10mm} (k_{\rm B}T \ll \hbar \Delta_{\rm eff}). 
\end{array} \right.
\end{eqnarray}
The $I$-$V$ characteristics
in the $K=1/2$ case is shown in Fig.~\ref{Fig3}~(b)
for $k_{\rm B}T \ll \hbar \Delta_{\rm eff}$ by the solid curve, and 
for $k_{\rm B}T \gg \hbar \Delta_{\rm eff}$ by the dashed curve.
This $I$-$V$ characteristics are consistent with the ones
obtained in Fig.~\ref{Fig3}~(a) when we take $K=1/2$ in our present result.

\subsection{Discussion based on a simplified model}
\label{sec34}

The above results can be understood intuitively 
in terms of the Coulomb blockade and the Bloch oscillation
by a simple analysis based on the equation of motion
\begin{equation}
  \dot Q + \frac{V_{\rm x}}{R} = I,
\label{model1}
\end{equation}
which is a simplified version
of the one analyzed in Ref.~\citen{Zaikin90} in detail.
Here, $Q$ and $I$ are the charge in the junction
and the external current, respectively.
The second term in (\ref{model1})
denotes the normal current through the shunted resistance.
The voltage is calculated by $V_{\rm x}={\rm d}E(Q)/{\rm d}Q$,
where $E(Q)$ is the dispersion relation given by (\ref{dispeq}).
The kinetic energy of the particle
originates from the Coulomb energy, and
$C_{\rm eff}=e^2/\hbar \Delta_{\rm eff}$ plays 
a role of an effective capacitance.
The Josephson effects appear from the fact that the 
pseudo-momentum $Q$ is folded into the region $-e \le Q \le e$.
For the tight-binding limit $E_{\rm J}/E_{\rm C} \gg 1$,
it is allowed to focus only on the lowest band.
We note that (\ref{model1}) is justified only when $Q$ is 
a classical variable and quantum fluctuations are negligible.

By solving the equation (\ref{model1}), we find
that the junction properties change at a
critical current $I^* = 2\pi\hbar\Delta_{\rm eff}/e R$.
For $I<I^*$, the charge converges to a finite value ($\dot{Q}=0$). 
In this regime, a Cooper pair cannot tunnel
through the junction, because it costs 
additional Coulomb energy for the Cooper-pair to tunnel.
This behavior is easily understood
as the Coulomb blockade phenomena.

For the opposite case $I > I^*$, the charge $Q$ continues to 
increase until $Q$ becomes $e$, and then the value of $Q$
jumps to $-e$ by tunneling of a Cooper pair.
This process iterates permanently. In this regime,
there appears voltage oscillation
with the frequency $f = I/2e$ 
known as the Bloch oscillation.~\cite{Zaikin90}

\begin{figure}[tb]
\hfil
\epsfxsize=65mm
\epsfbox{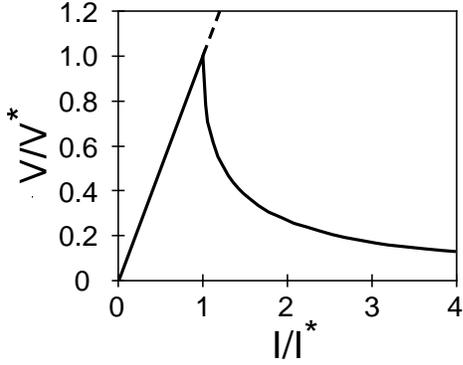}
\hfil
\caption{The $I$-$V$ characteristics calculated from
the simplified model (\ref{model1}). The current
and voltage is normalized by $I^* = 2\pi
\hbar \Delta_{\rm eff}/e R$
and $V^* =  2\pi \hbar \Delta_{\rm eff}/e$, 
respectively. The dashed line denotes 
the Ohmic response $V = IR$.}
\label{modelfig}
\end{figure}

From the simplified model (\ref{model1}),
the voltage averaged over the period $T=1/f$
is given as
\begin{equation}
V = \int_0^{T} \frac{{\rm d}t}{T} \, V_{\rm x},
\end{equation}
and is calculated as a function of $I$.
The result is shown in Fig.~\ref{modelfig}.
For $I < I^*$, the $I$-$V$ characteristics are
expressed exactly as $V=IR$, because no Cooper pair
can tunnel through the junction by the Coulomb blockade.
For $I > I^*$, the junction retains superconductivity 
by opening the channel of the Cooper-pair tunneling, and the
voltage is reduced. The whole shape of this $I$-$V$ 
characteristics is similar with the one obtained in the 
previous subsection. Particularly, the characteristic 
voltage $V^*$ and current $\eps^* = I^* \Phi_0$ are 
consistent with the one obtained in the last subsection 
($V^* \sim \hbar\Delta_{\rm eff}/e$, 
$\eps^* \sim K\hbar \Delta_{\rm eff}$). Thus, the two 
energy scales $\hbar \Delta_{\rm eff}$ and 
$K\hbar \Delta_{\rm eff}$ appear also in this simplified model.

We, however, should note that 
the simplified model does not reproduce
correct results at a quantitative level due to
the absence of quantum fluctuations of $Q$.
The differences can be pointed out as follows: (1)
the asymptotic behavior for $I \gg I^*$
of the simplified model is given by $V \propto 1/I$,
while the exact calculation gives $V \propto I^{2K-1}$,
(2) the exact solution for $K=1/2$ suggests that there is no
cusp in the $I$-$V$ characteristics for any $K$
in contrast to the result obtained from the simplified model,
and (3) the exact solution for $K=1/2$ also clarify that
even for small $I$ the superconductivity of the junction
is partially retained as seen in 
the deviation of the $I$-$V$ characteristics deviate
from the $V=IR$ line. 
Thus, it is clear that the simplified model 
cannot be used for the quantitative discussion. 

\section{Relation to the Kondo problem}
\label{sec4}

In this section, the dissipative tight-binding model
is discussed in relation to the Kondo problem. 
Although this discussion is not helpful in analyzing
unsolved problems about the
dissipative tight-binding model, it 
provides us with a new perspective,
which may contribute to a better understanding of the 
junction properties. 

First, we consider dissipative systems with a double-well potential.
When the systems are truncated to two-level systems, the 
Hamiltonian is obtained as
\begin{equation}
H = -\frac{\hbar \Delta}{2} \sigma_x +
\sum_{j} \left\{ \frac{p_j^2}{2m_j}
+ \frac{m_j \omega_j^2}{2} 
\left(x_j - \frac{c_j}{m_j \omega_j^2} \frac{a}{2} \sigma_z 
\right)^2 \right\}, 
\label{CLHam2} 
\end{equation}
where $\sigma_x$ and $\sigma_z$ are the Pauli matrices. 
This model has been studied as the simplest model
describing destruction of quantum coherence by surrounding
medium.~\cite{Leggett87,Weiss93}

In this paper, the dissipative `multi-level' system is studied. 
One may regard this model
as an extension of the dissipative `two-level' system.
There, however, exists a critical difference between these
two models. In dissipative two-level systems, the dissipation
is introduced to the `gaped' system with a gap $\hbar \Delta$,
while in multi-level systems, to the system with
a `gapless' band with a bandwidth $2\hbar \Delta$.
Though these two models show essentially the
same behavior at high temperature,
the low-temperature properties are quite different. 
Actually, density of states for dissipative two-level
systems has no characteristic feature at the energy scale
$K\hbar \Delta_{\rm eff}$ in contrast with the multi-level
systems.~\cite{Goerlich88}
The energy scale $K\hbar \Delta_{\rm eff}$ appears
in the dissipative two-level systems 
only as a characteristic relaxation time
in the correlation function and in the evolution of the expectation
value $P(t)=\langle \sigma_z (t) \rangle$.~\cite{Weiss89} 

Even in this dissipative two-level model,
it is a nontrivial task to study 
the system properties such as thermodynamics
and real-time correlation functions.
Actually, it has been proven that the low-energy behavior
of this dissipative two-level model is equivalent to
the Kondo model with anisotropic
exchange couplings:~\cite{Guinea85} 
\begin{equation}
H_{\rm K} = \sum_{k,\sigma} \eps_{k} c^{\dagger}_{k,\sigma}
c_{k,\sigma} + J_{\|} S_z s_z + J_{\bot} (S_x s_x + S_y s_y).
\end{equation}
Here, $c_{k,\sigma}$ ($c^{\dagger}_{k,\sigma}$) is
an annihilation (creation) operator of a conduction electron,
$S_i$ is the local spin operator, and $s_i$ is the spin operator
of the conduction electrons at the impurity site.
The equivalence between the Kondo model and the 
dissipative two-level system is obtained by identifying
the coupling constants $J_{\|}$ and $J_{\bot}$
with~\cite{Yuval70}
\begin{eqnarray}
\frac{\Delta}{\omega_c} &=& \rho J_{\bot} \cos^2 
\frac{\pi\rho J_{\|}}{4}, \label{map1} \\
K &=& \left( 1-\frac{\rho J_{\|}}{2} \right)^2,
\label{map2}
\end{eqnarray}
where $\rho$ is the density of states at the Fermi
surface. Unfortunately, the weak damping region $K \ll 1$
cannot be related to the Kondo problem exactly, since the above
relations become valid only in the limit 
$\rho J_{\|}, \rho J_{\bot} \ll 1$.~\cite{Yuval70}

The properties of
the dissipative two-level model can be understood in terms of
the Kondo problem as follows.~\cite{Leggett87}

In the dissipative two-level systems, the effective hopping
amplitude $\Delta_{\rm eff}$
determines the energy scale of the system.
At high temperatures $k_{\rm B}T \gg \hbar \Delta_{\rm eff}$,
the coherence of the system is destroyed by dissipation,
and the energy levels become degenerate. The particle
is localized at one well, and hops to the other
incoherently.
At low temperatures $k_{\rm B}T \ll \hbar \Delta_{\rm eff}$,
the coherence of the system
is retained, and the energy splitting is given by
$\hbar \Delta_{\rm eff}$. 

In the (anisotropic) Kondo model, the Kondo temperature $T_{\rm K}$
determines the energy scale of the system. At high 
temperatures $T \gg T_{\rm K}$, the local impurity spin 
fluctuates thermally, and the magnetic susceptibility 
of the local spin $\chi$ follows the Curie law. 
At low temperatures $T \ll T_{\rm K}$,
the local impurity spin forms a singlet state
coupled with conduction electrons, and $\chi$ becomes
independent of the temperature.

In the both cases, at a characteristic temperature
($\hbar \Delta_{\rm eff}$ or $k_{\rm B}T_{\rm K}$), the behavior of
the system changes drastically. This common feature 
actually comes from the equivalence of these two 
models.~\cite{Guinea85,Yuval70}

From this viewpoint, let us come back to the results obtained in the
previous section. In the dissipative multi-level model,
there are two different energy scales, which are seen clearly
in the calculation for $K \ll 1$. One is the effective bandwidth 
$2\hbar \Delta_{\rm eff}$, which can be related to the Kondo
temperature $T_{\rm K}$. This energy scale appears 
as a crossover temperature in the DC mobility, and
as a characteristic voltage $V \sim \hbar \Delta_{\rm eff}/2e$.
The other is the new energy scale $\hbar \gamma \sim K \hbar
\Delta_{\rm eff}$, which can be seen
in the density of states $D(\omega)$ (see Fig.~\ref{disp}),
and in the optical mobility $\mu(\omega)$. 
This energy scale is characteristic
of the multi-level model, in which 
the dissipation is introduced to a {\it gapless} system.
This situation cannot be realized in the Kondo problem.

In this sense, the resistance-shunted
Josephson junction system offers
a new type of nontrivial strongly-correlated 
problems. Though it has some similarity, it also has
an essential difference from the Kondo problem. 
If properly designed, Josephson
devices may show such a novel phenomenon originating from
strongly correlation effects.

\section{Summary}
\label{sec5}

In this paper, thermodynamics and linear responses
for the limiting region ($K \ll 1$ and $E_{\rm J}/E_{\rm C}
\gg 1$) have been studied analytically for all temperatures.
The nonlinear $I$-$V$ characteristics are also discussed
from the available results. 

The density of low-energy states 
($\omega < K\Delta_{\rm eff}$) is strongly modified.
This modification indicates that the dispersion relation
of relevant excitations is strongly altered at the band bottom.
The DC mobility $\mu_{\rm DC}(T) = V/I$ behaves as $T^{2K-2}$
at high temperatures, while it saturates to $R$ 
below a crossover temperature $T\sim \hbar \Delta_{\rm eff}
/k_{\rm B}$. At zero temperature,
the optical responses follow a Lorentzian form
with a width $\gamma = 4 \pi K \Delta_{\rm eff}$
at low frequencies, while they behave 
as $\omega^{2K-2}$ at high frequencies.
Above a crossover temperature 
$T \sim \hbar \Delta_{\rm eff}/k_{\rm B}$,
the optical mobility becomes temperature-dependent
on the low-frequency side. The nonlinear
$I$-$V$ characteristics for $0 < K < 1$
change at a crossover current $I^* \sim K \hbar
\Delta_{\rm eff}/\Phi_0$.
For $I < I^*$, the system behaves like an insulator
following the ohmic law $V=IR$, while for $I > I^*$
the junction retains the superconductivity, and behaves as $V\sim 
I^{2K-1}$. These features are reproduced qualitatively
by a simplified model. Among others, this system shows two
different energy scales. One is the effective bandwidth
$\hbar \Delta_{\rm eff}$, and the other is 
$K\hbar \Delta_{\rm eff}$. The latter energy scale
appears as the consequence of strong coupling to the
dissipation and is seen in the density of states, optical 
mobility and $I$-$V$ characteristics.
The Josephson devices can be
recognized as novel strongly-correlated systems.

In this paper, the junction properties have been studied
mainly in the weak damping region $K \ll 1$ and
the tight-binding limit $E_{\rm J}/E_{\rm C} \gg 1$.
These two conditions can be
realized easily in experimental situations.
In the tight-binding region, however,
the effective bandwidth $\hbar \Delta_{\rm eff}$
must be small (typically $\hbar \Delta_{\rm eff}/k_{\rm B}
< 1 \, {\rm [mK]}$). Then it becomes difficult to lower
the temperature below $\hbar \Delta_{\rm eff}/k_{\rm B}$. 
For experimental feasibility, the region $E_{\rm J}
\sim E_{\rm C}$ would be more fruitful, where
the qualitatively similar feature predicted
in this paper can be observed. Actually, also in the opposite
limit $E_{\rm J}/E_{\rm C} \ll 1$, two different energy
scales appear: the effective band width $\Delta = e^2/2C$
determines the characteristic temperature of the linear
mobility, while the damping energy $\gamma \sim K \Delta$
determines the characteristic current $\eps = I \Phi_0$
of the nonlinear mobility.~\cite{Fisher85}
The analyses in this paper may contribute
to the understanding of quantum effects in the Josephson junctions
also in those parameter regions.

The authors thank G. Sch\"on for useful suggestion.
T.K. is supported by Research Fellowship of 
Japan Society for the Promotion of Science 
for Young Scientists. This study was supported by a Grant-in-Aid
for Scientific Research from the Japanese Ministry of Education,
Science, Sports and Culture. This study was also supported
by a JSPS project
`Research for the Future' (JSPS-RFTF 97P01103).

\appendix

\section{Validity of the Debye-H\"uckel Approximation}
\label{app1}
The concept of this approximation has first been
considered by Debye and H\"uckel.~\cite{Debye23}
Later, Mayer has shown that the Debye-H\"uckel theory 
can be reproduced from the ring approximation
in the classical cluster expansion.~\cite{Mayer50}
Since then, the corrections to the Debye-H\"uckel approximation
have been studied for the problem of interacting plasma.

Let us consider the 3D classical Coulomb gas with the 
particle density $n$. In the Debye-H\"uckel approximation, 
the two-body distribution function is calculated as
\begin{equation}
\rho_2(r) = n^2 \exp\left(-\frac{\phi_{\rm D}(r)}{T} \right).
\end{equation}
Here, $T$ is the temperature, and $\phi_{\rm D}(r)$ is the 
screened potential
\begin{equation}
\phi_{\rm D}(r) = e^2 \frac{e^{-\kappa_{\rm D} r}}{r},
\end{equation}
where $1/\kappa_{\rm D}$ is the Debye screening length given by
\begin{equation}
\kappa_{\rm D} = \left(\frac{4\pi n e^2}{T}\right)^{1/2}.
\end{equation}
Note that the two-body distribution function can be written 
in a simple form as 
\begin{equation}
\rho_2(r) = n^2 \exp \left(-\eps \frac{e^{-x}}{x} \right),
\end{equation}
where $x =\kappa_{\rm D} r$ is the dimensionless spatial coordinate, and 
$\eps = e^2 \kappa_{\rm D}/T$ is the plasma parameter.
Beyond the ring approximation, it is possible to
sum the graphs up to the order of $\eps^2$ in the cluster 
expansion.~\cite{Ishihara69a,Ishihara69b} The result
is given as
\begin{equation}
\rho_2(r) = n^2 \exp \left(-\eps \frac{e^{-x}}{x}-\eps^2 f(x)\right),
\label{twobody}
\end{equation}
where $f(x)$ is a complex function of $x$. 
From this result, if the series of $\eps$ is convergent,
then the ring approximation is justified for $\eps \ll 1$.
Also the results by the Monte Carlo 
simulations~\cite{Slattery80,Slattery82} have supported
Eq.~(\ref{twobody}). Thus, the Debye-H\"uckel approximation 
is thought to be valid in the weak coupling limit $\eps \ll 1$ for
the 3D classical Coulomb gas.

One may suspect that the Debye-H\"ckel theory is not 
applicable to one-dimensional systems.
Here, let us discuss the 1D classical Coulomb gas
described by the Hamiltonian
\begin{equation}
H= \frac12 \sum_{i} p_i^2 - 2\pi e^2 \sum_{i<j} \sigma_i
\sigma_j | q_i - q_j |,
\end{equation}
where $p_j$, $q_j$, and $\sigma_j= \pm 1$ are the 
momentum, position, and sign of the $j$-th charged particle.
Among $2N$ particles, $N$ particles are charged positively and
$N$ negatively. Fortunately, this model can be solved
exactly.~\cite{Lenard61} We take the dimensionless
parameter as $\gamma = P/2\pi e^2$, where $P$ is the pressure. 
In this 1D model, the system behaves as ideal gas
in the high-pressure limit $\gamma \rightarrow \infty$
on the contrary to the 3D case.
The energy per particle $u$ is calculated for $\gamma \gg 1$
in the expansion form as
\begin{equation}
u = \frac{T}{2} + \frac{T}{2} \left\{
1 + \frac{1}{(2\gamma)^{1/2}} - \frac{1}{8\gamma} 
+ \frac{1}{128\sqrt{2} \gamma^{3/2}} + \cdots \right\}.
\label{1Dexact}
\end{equation}
It can easily be checked that the Debye-H\"uckel approximation
gives the exact result (\ref{1Dexact}) up to the order 
of $1/\gamma^{1/2}$. Thus, this approximation is also
applicable to the 1D classical Coulomb gas for the weak coupling 
region $\gamma \gg 1$.

From the above discussions, the Debye-H\"uckel approximation is 
thought to be applicable to general classical plasma models
in the condition 
\begin{equation}
1/\kappa_{\rm D} \gg \bar{l},
\label{condDH}
\end{equation}
where $\bar{l}$ is the averaged particle distance.
This condition is in fact reasonable from the physical viewpoint
because the Debye-H\"uckel approximation, an effective field
approach, is expected to become reliable when many particles are
interacting each other within the screening radius.
To clarify further the validity
of the ring approximation, other methods such as 
Monte Carlo simulations would be helpful.

To study the dissipative tight-binding model,
1D classical particles interacting with the potential 
$\phi(\tau)$ in the $\tau$ direction have been considered.
The potential is formulated as
\begin{eqnarray}
& & \phi(\tau) = \frac{1}{\hbar \beta} 
\sum_{\omega_m} \phi({\rm i}\omega_m)
e^{-{\rm i}\omega_m \tau},  \\
& & \phi({\rm i}\omega_m) = -\frac{2\pi K}{|\omega_m|},
\end{eqnarray}
for $0< |\omega_m| \ll \omega_{\rm c}$, 
while the potential vanishes for $\omega_m > \omega_{\rm c}$. 
The condition (\ref{condDH}) is evaluated 
for the dissipative tight-binding model as follows.
In the Debye-H\"uckel approximation, the screened potential 
$\vphi({\rm i}\omega_m)$ is calculated as
\begin{equation}
\vphi({\rm i}\omega_m) = \frac{\phi({\rm i}\omega_m)}
{1-n \phi({\rm i}\omega_m)/\hbar \beta}.
\label{screenedpot2}
\end{equation}
The screening effects appears when the second term
of the denominator in (\ref{screenedpot2}) is dominant.
Hence, the Debye screening length $1/\kappa_{\rm D}$ 
is determined from $n \phi(\omega_m=\kappa_D)/\hbar \beta \sim 1$,
and the condition (\ref{condDH}) corresponds to
the weak-damping condition $K \ll 1$. Thus, we expect
the ring approximation gives reliable results in the weak-damping
region.

It can be proven that the Debye-H\"uckel theory
reproduces correct results 
in the high temperature and/or strong damping region.~\cite{Kato98}
The Debye-H\"uckel approximation also
reproduces the zero-temperature mobility, which is
obtained by the renormalization group 
analysis as~\cite{Weiss93,Fisher85}
\begin{equation}
\mu_{\rm DC}(T=0) = R.
\end{equation}
Further, in the continuum limit
($K\rightarrow 0$, $\Delta \rightarrow \infty$ 
with $\gamma = 4\pi K \Delta$ fixed), 
it can be proven that the Debye-H\"uckel
approximation gives corrects results
for the optical mobility.~\cite{Kato98}
Also for thermodynamics, it can be checked that
the energy calculated in the Debye-H\"uckel approximation
\begin{equation}
E = {\rm const.} + \frac{\hbar \gamma}{2\pi}
\left[ \log\left(\frac{\hbar \beta \gamma}{2\pi}\right)
- \psi\left(\frac{\hbar \beta \gamma}{2\pi}\right)
-\frac{\pi}{\hbar \beta \gamma} \right]
\end{equation}
agrees with the energy of a dissipative Brownian
particle obtained in the continuum limit as~\cite{Weiss93}
\begin{equation}
\langle \frac{p^2}{2M} \rangle = \frac{\hbar}{2\pi}
\int_0^{\infty} {\rm d}\omega \frac{\omega \gamma}
{\omega^2 + \gamma^2} \coth(\omega \hbar \beta /2).
\end{equation}
These provide further supports for the reliability of the
Debye-H\"uckel theory in this problem.


\begin{thebibliography}{99}
\bibitem{footnote}The density of states in
the continuum limit
has not been calculated in Ref.~\citen{Kato98}.
\end{thebibliography}
\end{document}